\documentclass[aps,prd,preprintnumbers,nofootinbib,showpacs]{revtex4}
\usepackage{graphicx}
\usepackage{amsfonts}
\usepackage{amssymb,amsmath}
\usepackage{color}
\usepackage[colorlinks=true,linkcolor=blue,citecolor=blue]{hyperref}

\newcommand{\Phinu}{{\cal R}}
\newcommand{\Psinu}{{\cal S}}
\newcommand{\Qnu}{{\cal T}}

\newcommand{\simgt}{\lower.5ex\hbox{$\; \buildrel > \over \sim \;$}}
\newcommand{\simlt}{\lower.5ex\hbox{$\; \buildrel < \over \sim \;$}}

\begin{document}

\begin{center}
{\Large \bf
Bispectrum of cosmological density perturbations 
in the most general second-order scalar-tensor theory
}

\vskip .45in

{
Yuichiro Takushima${}^1$, Ayumu Terukina${}^1$, and Kazuhiro Yamamoto${}^{1,2}$
}

\vskip .45in

{\em
${}^1$Department of Physical Sciences, Hiroshima University, 
Higashi-hiroshima, Kagamiyama 1-3-1, 739-8526, Japan
\\
${}^2$Hiroshima Astrophysical Science Center, Hiroshima University, Higashi-Hiroshima,
Kagamiyama 1-3-1, 739-8526, Japan
}

\end{center}

\begin{abstract}
We study the bispectrum of the matter density perturbations induced
by the large scale structure formation in the most general second-order 
scalar-tensor theory that may possess the Vainshtein mechanism as 
a screening mechanism. 
On the basis of the standard perturbation theory, we derive 
the bispectrum being expressed by a kernel of the second order 
of the density perturbations. 
We find that the kernel at the leading order is characterized by one parameter, 
which is determined by the solutions of the linear density perturbations, 
the Hubble parameter and the other function specifying nonlinear interactions. 
This does not allow for varied behavior in the bispectrum of the matter 
density perturbations in the most general second-order scalar-tensor 
theory equipped with the Vainshtein mechanism.
We exemplify the typical behavior of the bispectrum in a kinetic gravity 
braiding model. 
\end{abstract}

\maketitle

\setcounter{page}{1}
\section{Introduction} 
\label{sec:intro}
Modified gravity models attract interests of researchers as an alternative to 
explain the cosmic accelerated expansion of the universe without introducing 
the cosmological constant \cite{HuSawicki,Starobinsky,Tsujikawafr,Nojirib,DGP2,Song,Maartens,KM,MassiveG,MassiveG2,MassiveG3,HA}.
The most general second-order scalar-tensor theory was constructed by Horndeski
\cite{Horndeski} for the first time, and it was rediscovered in \cite{DGSZ}
as a generalization of the galileon theories \cite{Koba,GC,SAUGC,ELCP,CEGH,GGC,GBDT,DPGMGT,CG,GGIR,MGALG,CCGF,OCG,CSTOG,FeliceTsujikawa,Deffayet,DDEF,GALMG,RFFGM,SSSDBI,DBIGR}.
In addition to the possibility of constructing cosmological models with an 
accelerated expansion, it possesses the following interesting features. 
The equation of motion are the second order differential equation. 
Then, an additional degree of freedom is not introduced, which is 
advantageous to avoid the appearance of ghosts. 
Furthermore, the galileon theory is endowed with the Vainshtein 
mechanism \cite{GALMG}, which is a screening mechanism 
useful to evade the local gravity constraints. 
In the most general second-order scalar-tensor theory, the Vainshtein 
mechanism may work depending on the model parameters (e.g., \cite{KKY,Kase,Narikawa}). 

The results of the Planck satellite have shown that the primordial perturbations obey
almost the Gaussian statistics \cite{Planckresults}. Even if the initial perturbations 
were completely Gaussian, the non-Gaussian nature in the density 
perturbations is induced in the large scale structure formation through the nonlinear 
fluid equations under the influence of the gravitational force. 
The bispectrum is often used to characterize the nonlinear and non-Gaussian 
nature in the density perturbations (e.g., \cite{Bispectrum1,Bispectrum2,Bispectrum3,BMR1,BMR2}). 
Recently, bispectrum and nonlinear features in the structure formation in the 
galileon models have been investigated \cite{KTH,Barreira,BLi,Wyman,Bellini,Emilio,kgb}.
In the present paper, we focus our investigation on the bispectrum
in the most general second-order scalar-tensor theory in order to 
illuminate characteristic features in a wide class of modified 
gravity models, regarding it as an effective theory. 
An advantage of such a general theory is that we can discuss general 
features of a wide class of modified gravity models, 
which is useful to forecast their detectability in future large surveys.

In the present paper, we consider the bispectrum in the matter 
density perturbations which is induced in the large scale structure 
formation after the matter dominated epoch. 
We present an expression of the bispectrum in the most general 
second-order scalar-tensor theory based on the standard density 
perturbation theory, which is written in term of 
a kernel of the second order of perturbations. We find that 
the kernel is characterized by only {\it one} parameter,
which is determined by the solutions of the linear density perturbations, 
the Hubble parameter, and the other function of the background universe
that describes the nonlinear interactions. 
This paper is organized as follows: 
In section 2, we apply the standard perturbation theory to the 
most general second-order scalar-tensor theory that may possess 
the Vainshtein mechanism, and find the solution of the second-order 
of density perturbations. 
In section 3, we present the expression of the bispectrum of the 
density perturbations, and investigate the influence of the modification 
of gravity. The results are applied to a simple kinetic gravity braiding
model in section 4. Section 5 is devoted to summary and conclusions.

\section{Formulation} 
\label{sec:fm}
We consider the most general second-order scalar-tensor theory
on the expanding universe background.
The action is given by  
\begin{eqnarray}
S=\int d^4x\sqrt{-g}\left({\cal L}_{\rm GG}+{\cal L}_{\rm m}\right),\label{action}
\label{action}
\end{eqnarray}
where we defined 
\begin{eqnarray}
{\cal L}_{\rm GG} &=& K(\phi, X)-G_3(\phi, X)\Box\phi
+G_4(\phi, X)R+G_{4X}\left[(\Box\phi)^2-(\nabla_\mu\nabla_\nu\phi)^2\right]
\nonumber\\
&&+G_5(\phi, X)G_{\mu\nu}\nabla^\mu\nabla^\nu\phi
-\frac{1}{6}G_{5X}\bigl[(\Box\phi)^3
-3\Box\phi(\nabla_\mu\nabla_\nu\phi)^2+
2(\nabla_\mu\nabla_\nu\phi)^3\bigr],\label{GG}
\end{eqnarray}
with four arbitrary functions,
$K, G_3, G_4,$ and $G_5$, of $\phi$ and $X:=-(\partial\phi)^2/2$,
$G_{iX}$ stands for $\partial G_i/\partial X$, 
$R$ is the Ricci scalar, $G_{\mu\nu}$ is the Einstein tensor, 
and ${\cal L}_{\rm m}$ is the matter Lagrangian, which is assumed to be 
minimally coupled to gravity. 
This theory is found in \cite{DGSZ} as a generalization of the galileon theory, 
but the equivalence with the Horndeski's theory is shown in \cite{Koba}.
We consider a spatially  flat expanding universe and the metric 
perturbations in the Newtonian gauge, whose line element is written 
as
\begin{eqnarray}
ds^2=-(1+2\Phi)dt^2+a^2(1-2\Psi)d\mathbf{x}^2.
\end{eqnarray}
We define the scalar field with perturbations by 
\begin{eqnarray}
\phi&\to&\phi(t)+\delta\phi(t, \mathbf{x}),
\end{eqnarray}
with which we introduce $Q:=H{\delta\phi}/{\dot\phi}$.

We consider the case that the Vainshtein mechanism may work as a 
screening mechanism. 
The basic equations for the cosmological density perturbations are 
derived in Ref.~\cite{KKY}. Here we briefly review the method and 
the results (see \cite{KKY} for details). 
The basic equations of the gravitational and scalar fields are derived 
on the basis of the quasi-static approximation
of the subhorizon scales. 
The models that the Vainshtein mechanism may work can be found as follows. 
The equations are derived by
keeping the leading terms schematically written as $(\partial\partial Y)^n$, 
with $n\geq1$, where $\partial$ denotes a spatial derivative and $Y$
does any of $\Phi$, $\Psi$ or $Q$.
Such terms make a leading contribution of the order $(L_{\rm H}^2\partial\partial Y)^n$, 
where $L_{\rm H}$ is a typical horizon length scale. 
According to Ref.~\cite{KKY}, from the gravitational field equation, 
we have
\begin{eqnarray}
&&\nabla^2\left({\cal F}_T\Psi-{\cal G}_T\Phi-A_1 Q\right)
=\frac{B_1}{2a^2H^2}{\cal Q}^{(2)}
+\frac{B_3}{a^2H^2}\left(
\nabla^2\Phi\nabla^2Q-\partial_i\partial_j\Phi\partial^i\partial^j Q
\right),
\label{trlseq}
\\
&&{\cal G}_T\nabla^2\Psi
=\frac{a^2}{2}\rho_{\rm m}\delta
-A_2 \nabla^2 Q
-\frac{B_2}{2a^2H^2} {\cal Q}^{(2)}
-\frac{B_3}{a^2H^2}\left(\nabla^2\Psi\nabla^2Q
-\partial_i\partial_j\Psi\partial^i\partial^jQ\right)
-\frac{C_1}{3a^4H^4}{\cal Q}^{(3)},\label{00eq}
\end{eqnarray}
where $\rho_{\rm m}$ is the matter density, $\delta$ is the matter 
density contrast, and we defined
\begin{eqnarray}
&&{\cal Q}^{(2)}:=\left(\nabla^2Q\right)^2-\left(\partial_i\partial_j Q\right)^2,
\\
&&{\cal Q}^{(3)}:=\left(\nabla^2 Q\right)^3
-3\nabla^2Q\left(\partial_i\partial_jQ\right)^2
+2\left(\partial_i\partial_jQ\right)^3.
\end{eqnarray}
{}From the equation of motion of the scalar field, we have
\begin{eqnarray}
&&A_0\nabla^2Q
-A_1\nabla^2\Psi
-A_2\nabla^2\Phi+\frac{B_0}{a^2H^2}{\cal Q}^{(2)}
-\frac{B_1}{a^2H^2}
\left(\nabla^2\Psi\nabla^2Q-\partial_i\partial_j\Psi\partial^i\partial^jQ\right)
\nonumber\\&&
-\frac{B_2}{a^2H^2}
\left(\nabla^2\Phi\nabla^2Q-\partial_i\partial_j\Phi\partial^i\partial^jQ\right)
-\frac{B_3}{a^2H^2}
\left(\nabla^2\Phi\nabla^2\Psi -
\partial_i\partial_j\Phi\partial^i\partial^j\Psi \right)
\nonumber\\&&
-\frac{C_0}{a^4H^4}{\cal Q}^{(3)}
-\frac{C_1}{a^4H^4}{\cal U}^{(3)}= 0,
\label{seom}
\end{eqnarray}
where we defined
\begin{eqnarray}
{\cal U}^{(3)}&:=&
{\cal Q}^{(2)}\nabla^2\Phi
-2\nabla^2Q\partial_i\partial_jQ\partial^i\partial^j\Phi
+2\partial_i\partial_jQ\partial^j\partial^kQ\partial_k\partial^i\Phi.
\end{eqnarray}
The coefficients such as ${\cal F}_T$,
$A_1$, $B_1$, etc.,  that appear in the field equations here and
hereafter are defined in Appendix A. $A_i,B_i,C_i$ are the 
coefficients of the linear, quadratic and cubic terms of
$\Psi$, $\Phi$, $Q$, respectively. 

Equations for the matter density contrast $\delta$ and the velocity field 
$u^i$ are given by 
\begin{eqnarray}
&&{\partial \delta(t,{\bf x})\over \partial t}
+{1\over a}\partial_i[(1+\delta(t,{\bf x})) u^i(t,{\bf x})]=0,
\label{continue}
\\
&&{\partial u^i(t,{\bf x})
\over \partial t}+{\dot a\over a}u^i(t,{\bf x})
+{1\over a}u^j(t,{\bf x})\partial_ju^i(t,{\bf x})
=-{1\over a}\partial^i\Phi(t,{\bf x}), 
\label{Euler}
\end{eqnarray}
where the dot denotes the differentiation with respect to $t$. 
The effect of the gravity comes through the gravitational potential $\Phi$, 
which is determined by the above equations 
(\ref{trlseq}), (\ref{00eq}) and (\ref{seom}). 
Here, we only consider the scalar mode of the density perturbations, 
then we introduce a scalar function by $\theta\equiv \nabla {\bf u}/(aH)$. 
Now we define the Fourier expansion as for the quantities 
$\delta$ and $\theta$,
\begin{eqnarray}
&&\delta(t,{\bf x})={1\over (2\pi)^3}\int d^3p\delta(t,{\bf p})e^{i{\bf p}\cdot{\bf x}},
\label{deftildedelta}
\\
&&u^j(t,{\bf x})={1\over (2\pi)^3}\int d^3p {-ip^j \over p^2}aH\theta(t,{\bf p})e^{i{\bf p}\cdot{\bf x}}.
\end{eqnarray}
The Fourier expansion for $\Phi$, $\Psi$ and $Q$ 
is defined in the similar way to (\ref{deftildedelta}).
Then, (\ref{trlseq}) and (\ref{00eq}) yield
\begin{eqnarray}
&&-p^2\left({\cal F}_T\Psi(t,{\bf p})-{\cal G}_T\Phi(t,{\bf p})-A_1 Q(t,{\bf p})\right)
=\frac{B_1}{2a^2H^2}\Gamma[t,{\bf p};Q,Q]
+\frac{B_3}{a^2H^2}\Gamma[t,{\bf p};Q,\Phi]
\label{se14}\\
&&-p^2({\cal G}_T\Psi(t,{\bf p})+A_2 Q(t,{\bf p}))
-\frac{a^2}{2}\rho_{\rm m}\delta(t,{\bf p})
=-\frac{B_2}{2a^2H^2} \Gamma[t,{\bf p};Q,Q]
-\frac{B_3}{a^2H^2}\Gamma[t,{\bf p};Q,\Psi]
\nonumber\\&&
~~~~~~~-\frac{C_1}{3a^4H^4}
{1\over (2\pi)^6}\int d{\bf k}_1 d{\bf k}_2d{\bf k}_3\delta^{(3)}({\bf k}_1+{\bf k}_2+{\bf k}_3-{\bf p})
\nonumber\\&&
\hspace{0cm}
~~~~~~~\times \biggl[
-k_1^2k_2^2k_3^2+3k_1^2({\bf k}_2\cdot{\bf k}_3)^2-2({\bf k}_1\cdot{\bf k}_2)({\bf k}_2\cdot{\bf k}_3)({\bf k}_3\cdot{\bf k}_1)\biggr]Q(t,{\bf k}_1)Q(t,{\bf k}_2)Q(t,{\bf k}_3),
\label{se15}
\end{eqnarray}
where we defined
\begin{eqnarray}
\Gamma[t,{\bf p};Y,Z]={1\over (2\pi)^3}\int d{\bf k}_1 d{\bf k}_2\delta^{(3)}({\bf k}_1+{\bf k}_2-{\bf p})
\left(k_1^2k_2^2-({\bf k}_1\cdot{\bf k}_2)^2\right)Y(t,{\bf k}_1)Z(t,{\bf k}_2),
\end{eqnarray}
where $Y$ and $Z$ denote any of $Q$, $\Phi$, or $\Psi$.
Eq.~(\ref{seom}) leads to
\begin{eqnarray}
&&-p^2(A_0Q(t,{\bf p})
-A_1\Psi(t,{\bf p})
-A_2\Phi(t,{\bf p}))
\nonumber\\
&&~~~~=-\frac{B_0}{a^2H^2}\Gamma[t,{\bf p};Q,Q]
+\frac{B_1}{a^2H^2}\Gamma[t,{\bf p};Q,\Psi]
+\frac{B_2}{a^2H^2}\Gamma[t,{\bf p};Q,\Phi]
+\frac{B_3}{a^2H^2}\Gamma[t,{\bf p};\Psi,\Phi]
\nonumber\\
&&~~~~
+\frac{C_0}{a^4H^4}
{1\over (2\pi)^6}\int d{\bf k}_1 d{\bf k}_2d{\bf k}_3\delta^{(3)}({\bf k}_1+{\bf k}_2+{\bf k}_3-{\bf p})
\biggl[
-k_1^2k_2^2k_3^2+3k_1^2({\bf k}_2\cdot{\bf k}_3)^2
\nonumber\\&&
~~~~\hspace{0.1cm}
-2({\bf k}_1\cdot{\bf k}_2)({\bf k}_2\cdot{\bf k}_3)({\bf k}_3\cdot{\bf k}_1)\biggr]Q(t,{\bf k}_1)Q(t,{\bf k}_2)Q(t,{\bf k}_3)
\nonumber\\
&&
~~~~+\frac{C_1}{a^4H^4}
{1\over (2\pi)^6}\int d{\bf k}_1 d{\bf k}_2d{\bf k}_3\delta^{(3)}({\bf k}_1+{\bf k}_2+{\bf k}_3-{\bf p})
\biggl[
-k_1^2k_2^2k_3^2+({\bf k}_1\cdot{\bf k}_2)^2k_3^2
\nonumber\\&&
~~~~\hspace{0.1cm}
+2k_1^2({\bf k}_2\cdot{\bf k}_3)^2
-2({\bf k}_1\cdot{\bf k}_2)({\bf k}_2\cdot{\bf k}_3)({\bf k}_3\cdot{\bf k}_1)\biggr]
Q(t,{\bf k}_1)Q(t,{\bf k}_2)\Phi(t,{\bf k}_3).
\label{seomf}
\end{eqnarray}
Equations (\ref{continue}) and (\ref{Euler}) are rephrased as
\begin{eqnarray}
&&{1\over H}{\partial \delta(t,{\bf p})\over \partial t}+\theta(t,{\bf p})
=-{1\over (2\pi)^3}\int d{\bf k}_1 d{\bf k}_2\delta^{(3)}({\bf k}_1+{\bf k}_2-{\bf p})
\left(1+{{\bf k}_1\cdot{\bf k}_2\over k_2^2}\right)\delta(t,{\bf k}_1)\theta(t,{\bf k}_2),
\label{se11}
\\
&&{1\over H}{\partial \theta(t,{\bf p})\over \partial t}+
\left(2+{\dot H\over H^2}\right)\theta(t,{\bf p})-{p^2\over a^2H^2}\Phi(t,{\bf p})
\nonumber\\
&&\hspace{1cm}
=-{1\over2}{1\over (2\pi)^3}\int d{\bf k}_1 d{\bf k}_2\delta^{(3)}({\bf k}_1+{\bf k}_2-{\bf p})
\left({({\bf k}_1\cdot {\bf k}_2)|{\bf k}_1+{\bf k}_2|^2\over k_1^2k_2^2}\right)
\theta(t,{\bf k}_1)\theta(t,{\bf k}_2).
\label{se12}
\end{eqnarray}

We find the solution in terms of the perturbative expansion, which can 
be written in a form
\begin{eqnarray}
Y(t,{\bf p}) &=& \sum_{n=1}Y_n(t,{\bf p}),
\end{eqnarray}
where $Y$ denotes $\delta,\theta,\Psi,\Phi$, or $Q$, and
$Y_n$ denotes the $n$-th order solution of the expansion. 

Now we start from the first order equations, 
which can be easily solved as follows. 
{}From equations (\ref{se14}), (\ref{se15}), and (\ref{seomf}), we have
\begin{eqnarray}
&&{\cal F}_T p^2 \Psi_1(t,{\bf p}) - {\cal G}_T p^2 \Phi_1(t,{\bf p}) 
- A_1 p^2Q_1(t,{\bf p}) =0,
\\
&&{\cal G}_T p^2\Psi_1(t,{\bf p}) + A_2 p^2 Q_1(t,{\bf p}) = 
- {a^2\over 2}\rho_{\rm m} \delta_1(t,{\bf p}),
\\
&&A_0 p^2 Q_1(t,{\bf p}) - A_1 p^2 \Psi_1(t,{\bf p}) - A_2 p^2 \Phi_1(t,{\bf p}) =0,
\end{eqnarray}
which give the solutions 
\begin{eqnarray}
\Phi_1(t,{\bf p})&=&-{a^2 \rho_{\rm m} {\cal R}(t)\over p^2 {\cal Z}(t)} \delta_1(t,{\bf p}),
\label{eqPhifo}
\\
\Psi_1(t,{\bf p})&=&-{a^2 \rho_{\rm m}{\cal S}(t)\over p^2 {\cal Z}(t)} \delta_1(t,{\bf p}),
\label{eqPsifo}
\\
Q_1(t,{\bf p})&=&-{a^2 \rho_{\rm m}{\cal T}(t)\over p^2{\cal Z}(t)} \delta_1(t,{\bf p}),
\label{eqQfo}
\end{eqnarray}
where we defined
\begin{eqnarray}
&&\Phinu(t) =A_0 {\cal F}_T - A_1^2,
\\
&&\Psinu(t)=A_0 {\cal G}_T + A_1 A_2,
\\
&&\Qnu(t) =A_1 {\cal G}_T + A_2 {\cal F}_T,
\\
&&{\cal Z}(t)=2 (A_0 {\cal G}_T^2 + 2 A_1 A_2 {\cal G}_T +A_2^2{\cal F}_T).
\end{eqnarray}
The first order equation of (\ref{se11}) is 
\begin{eqnarray}
\theta_1(t,{\bf p}) = - {1\over H}{\partial \delta_1(t,{\bf p}) \over \partial t}.
\label{se11fo}
\end{eqnarray}
Substituting (\ref{se11fo}) and (\ref{eqPhifo}) into the first order equation of 
(\ref{se12}), we have
\begin{eqnarray}
&&{\partial^2\delta_1 (t,{\bf p}) \over \partial t^2} + 2 H {\partial \delta_1(t,{\bf p}) \over \partial t} + L(t)\delta_1(t,{\bf p})=0,
\label{lineareq}
\end{eqnarray}
where we defined
\begin{eqnarray}
&&L(t)
= -{(A_0 {\cal F}_T - A_1^2)\rho_{\rm m} \over 2 (A_0 {\cal G}_T^2 + 2 A_1 A_2{\cal G}_T + A_2^2 {\cal F}_T)}.
\end{eqnarray}
This second rank differential equation has the growing mode solution $D_+(t)$ and the 
decaying mode solution $D_-(t)$. Neglecting the decaying mode solution, we 
write the first order solution,
\begin{eqnarray}
\delta_1(t,{\bf p}) = D_+(t)\delta_{\rm L}({\bf p}),
\label{soldelta1}
\end{eqnarray}
where $\delta_{\rm L}({\bf p})$ is a constant, which is determined by the 
initial density fluctuations. We assume that $\delta_{\rm L}({\bf p})$
obeys the Gaussian random statistics. Here we adopt the normalization $D_+(a)=a$
at $a\ll1$. The first order solutions for the other quantities can be 
expressed in terms of $\delta_1(t,{\bf p})$.

Then, we consider the second order equations of the perturbative expansion.
{}From (\ref{se14}), (\ref{se15}) and (\ref{seomf}), the second order equations 
are
\begin{eqnarray}
&&-p^2\left({\cal F}_T\Psi_2(t,{\bf p})-{\cal G}_T\Phi_2(t,{\bf p})-A_1 Q_2(t,{\bf p})\right)
=\frac{B_1}{2a^2H^2}\Gamma[t,{\bf p};Q_1,Q_1]
+\frac{B_3}{a^2H^2}\Gamma[t,{\bf p};Q_1,\Phi_1],
\label{se14b}
\\
&&-p^2({\cal G}_T\Psi_2(t,{\bf p})+A_2 Q_2(t,{\bf p})
=\frac{a^2}{2}\rho_{\rm m}\delta_2(t,{\bf p})
-\frac{B_2}{2a^2H^2} \Gamma[t,{\bf p};Q_1,Q_1]
-\frac{B_3}{a^2H^2}\Gamma[t,{\bf p};Q_1,\Psi_1],
\label{se15b}
\\
&&-p^2(A_0Q_2(t,{\bf p})
-A_1\Psi_2(t,{\bf p})
-A_2\Phi_2(t,{\bf p}))
\nonumber\\&&~~~~
=-\frac{B_0}{a^2H^2}\Gamma[t,{\bf p};Q_1,Q_1]
+\frac{B_1}{a^2H^2}\Gamma[t,{\bf p};Q_1,\Psi_1]
+\frac{B_2}{a^2H^2}\Gamma[t,{\bf p};Q_1,\Phi_1]
+\frac{B_3}{a^2H^2}\Gamma[t,{\bf p};\Psi_1,\Phi_1]
.\label{seomfb}
\end{eqnarray}
Using the first order solutions (\ref{eqPhifo}), (\ref{eqPsifo}), (\ref{eqQfo}), 
and (\ref{soldelta1}), the above equations are rephrased as
\begin{eqnarray}
&&\hspace{-1cm}-p^2\left({\cal F}_T\Psi_2(t,{\bf p})-{\cal G}_T\Phi_2(t,{\bf p})-A_1 Q_2(t,{\bf p})\right)
={D_+^2(t) a^2 \rho_{\rm m}^2 \over H^2 {\cal Z}^2(t)}\Bigl({1\over 2} B_1 \Qnu^2(t) + B_3 \Qnu(t)\Phinu(t)\Bigr){\cal W}_\gamma({\bf p}),
\\
&&\hspace{-1cm}-p^2({\cal G}_T\Psi_2(t,{\bf p}) + A_2 Q_2(t,{\bf p}))
=\frac{a^2}{2}\rho_{\rm m}\delta_2(t,{\bf p})
+{D_+^2(t) a^2 \rho_{\rm m}^2  \over H^2 {\cal Z}^2(t)}\Bigl(-{1\over 2} B_2 \Qnu^2(t) - B_3 \Qnu(t)\Psinu(t)\Bigr){\cal W}_\gamma({\bf p}),
\\
&&\hspace{-1cm}-p^2(A_0Q_2(t,{\bf p})-A_1\Psi_2(t,{\bf p})-A_2\Phi_2(t,{\bf p})) \nonumber\\
&&
= {D_+^2(t) a^2 \rho_{\rm m}^2 \over H^2 {\cal Z}^2(t)}\Bigl(-B_0 \Qnu^2(t) + B_1 \Psinu(t)\Qnu(t) + B_2\Phinu(t)\Qnu(t) + B_3 \Phinu(t)\Psinu(t)\Bigr){\cal W}_\gamma({\bf p}),
\end{eqnarray}
where we defined
\begin{eqnarray}
&&{\cal W}_\gamma({\bf p})= {1\over(2 \pi)^3}\int d{\bf k}_1 d{\bf k}_2\delta^{(3)}({\bf k}_1+{\bf k}_2-{\bf p})\gamma({\bf k}_1,{\bf k}_2)\delta_{\rm L}({\bf k}_1)\delta_{\rm L}({\bf k}_2),
\\
&&\gamma({\bf k}_1,{\bf k}_2) = 1-{({\bf k}_1\cdot{\bf k}_2)^2\over k_1^2k_2^2}.
\end{eqnarray}
These equations yield
\begin{eqnarray}
&&\hspace{-1cm}
\Phi_2(t,{\bf p}) =-{a^2 \rho_{\rm m}  \Phinu \over p^2 {\cal Z}} \delta_2(t,{\bf p})
-{D_+^2(t) a^2 \rho_{\rm m}^2{\cal W}_\gamma({\bf p})\over H^2p^2{\cal Z}^3}
\Bigl\{2 B_0 \Qnu^3 
- 3 B_1\Psinu \Qnu^2 -3 B_2\Phinu \Qnu^2 - 6 B_3\Phinu \Psinu \Qnu\Bigr\},
\label{solphi22}
\\
&&\hspace{-1cm}
\Psi_2(t,{\bf p}) =-{a^2 \rho_{\rm m} \Psinu \over p^2 {\cal Z}}  \delta_2(t,{\bf p})
-{D_+^2(t) a^2 \rho_{\rm m}^2{\cal W}_\gamma({\bf p})\over H^2p^2{\cal Z}^3}
\Bigl\{2 B_0 A_2{\cal G}_T \Qnu^2 
+  B_1(A_2^2 \Qnu^2 - 2 A_2 {\cal G}_T \Psinu \Qnu) 
\nonumber\\
&&~~~~~
-B_2(\Psinu \Qnu^2 - 2 A_2 {\cal G}_T \Phinu \Qnu)
-B_3(2 \Psinu^2 \Qnu - 2 A_2^2 \Phinu \Qnu + 2 A_2 {\cal G}_T \Phinu \Psinu)
\Bigr\},
\\
&&\hspace{-1cm}
Q_2(t,{\bf p}) =-{a^2 \rho_{\rm m}  \Qnu \over p^2 {\cal Z}}  \delta_2(t,{\bf p})
+{D_+^2(t) a^2 \rho_{\rm m}^2{\cal W}_\gamma({\bf p})\over H^2p^2{\cal Z}^3}\Bigl\{2 B_0 {\cal G}_T^2 {\cal T}^2 
+ B_1(A_2 {\cal G}_T \Qnu^2 - 2 {\cal G}_T^2 \Psinu\Qnu) 
\nonumber\\
&&~~~~~
+ B_2(\Qnu^3 - 2 {\cal G}_T^2 \Phinu \Qnu)
+B_3(2\Psinu \Qnu^2 + 2 A_2 {\cal G}_T \Phinu \Qnu - 2 {\cal G}_T^2 \Phinu \Psinu)
\Bigr\}.
\end{eqnarray}
The second order equations of (\ref{se11}) and (\ref{se12}) are
\begin{eqnarray}
&&{1\over H}{\partial \delta_2(t,{\bf p})\over \partial t} + \theta_2(t,{\bf p}) 
=-{1\over (2\pi)^3}\int d{\bf k}_1 d{\bf k}_2\delta^{(3)}({\bf k}_1+{\bf k}_2-{\bf p})
\alpha({\bf k}_1,{\bf k}_2)\delta_1(t,{\bf k}_1)\theta_1(t,{\bf k}_2),
\label{cont2}
\\
&&{1\over H}{\partial \theta_2(t,{\bf p})\over\partial t} + \left( 2 + {\dot{H}\over H^2}\right)\theta_2(t,{\bf p}) - {p^2\over a^2 H^2}\Phi_2(t,{\bf p})\nonumber\\
&&\hspace{3cm}
=-{1\over (2\pi)^3}\int d{\bf k}_1 d{\bf k}_2\delta^{(3)}({\bf k}_1+{\bf k}_2-{\bf p})
\beta({\bf k}_1,{\bf k}_2)
\theta_1(t,{\bf k}_1)\theta_1(t,{\bf k}_2),
\label{Euler2}
\end{eqnarray}
where we defined
\begin{eqnarray}
&&\alpha({\bf k}_1,{\bf k}_2) = 1 + {{\bf k}_1 \cdot {\bf k}_2\over k_1^2},
\\
&&\beta({\bf k}_1,{\bf k}_2) = {({\bf k}_1 \cdot {\bf k}_2) |{\bf k}_1 + {\bf k}_2 |^2\over 2 k_1^2 k_2^2}.
\end{eqnarray}
Combining (\ref{cont2}) and (\ref{Euler2}), and using the 
first order solution and (\ref{solphi22}), we have
\begin{eqnarray}
&&{\partial^2\delta_2 (t,{\bf p}) \over \partial t^2} + 2 H {\partial \delta_2(t,{\bf p}) \over \partial t} + L(t)\delta_2(t,{\bf p})= S_\delta(t,{\bf p}),
\label{eqdel1}
\end{eqnarray}
where we defined
\begin{eqnarray}
&&S_\delta(t,{\bf p})
=\left(\dot D_+^2(t)-L(t)D_+^2(t)\right){\cal W}_\alpha({\bf p}) 
+ \dot {D}_+^2 (t)
{\cal W}_\beta({\bf p}) +  {N}_{\gamma}(t)D_+^2(t){\cal W}_\gamma({\bf p}),
\label{eqdel2}
\\
&&{\cal W}_\alpha({\bf p}) = {1\over (2\pi)^3}\int d{\bf k}_1 d{\bf k}_2 \delta^{(3)}({\bf k}_1 + {\bf k}_2 - {\bf p})\alpha({\bf k}_1,{\bf k}_2)\delta_{\rm L}({\bf k}_1)\delta_{\rm L}({\bf k}_2), 
\\
&&{\cal W}_\beta({\bf p}) = {1\over (2\pi)^3}\int d{\bf k}_1 d{\bf k}_2 \delta^{(3)}({\bf k}_1 + {\bf k}_2 - {\bf p})\beta({\bf k}_1,{\bf k}_2)\delta_{\rm L}({\bf k}_1)\delta_{\rm L}({\bf k}_2),
\end{eqnarray}
and 
\begin{eqnarray}
&&N_\gamma (t) = {\rho_{\rm m}^2 \over H^2 {\cal Z}^3} \left(
2 B_0 \Qnu^3 - 3 B_1 \Psinu \Qnu^2 -3 B_2 \Phinu \Qnu^2 -6 B_3 \Phinu\Psinu \Qnu \right).
\end{eqnarray}
In deriving (\ref{eqdel2}), we used (\ref{lineareq}).
Because of the symmetry with respect to the interchange of ${\bf k}_1$ and ${\bf k}_2$, 
we redefine $\alpha({\bf k}_1,{\bf k}_2)$ as follows,
\begin{eqnarray}
\alpha({\bf k}_1,{\bf k}_2) &=& 1 + {
{\bf k}_1 \cdot {\bf k}_2 (k_1^2 + k_2^2)\over 2 k_1^2 k_2^2}.
\end{eqnarray}
Using the relation, 
\begin{eqnarray}
\beta({\bf k}_1,{\bf k}_2)=\alpha({\bf k}_1,{\bf k}_2)-\gamma({\bf k}_1,{\bf k}_2)
~~{\rm or}~~ {\cal W}_\beta({\bf p}) = {\cal W}_\alpha({\bf p}) - {\cal W}_\gamma({\bf p}), 
\end{eqnarray}
equation (\ref{eqdel2}) reduces to 
\begin{eqnarray}
 S_\delta(t,{\bf p})
&=&\left(2f^2H^2-L(t)\right)D_+^2(t){\cal W}_\alpha({\bf p}) 
+  \left({N}_{\gamma}(t)-f^2H^2\right)D_+^2(t){\cal W}_\gamma({\bf p}),
\label{eqdel3}
\end{eqnarray}
where we defined the growth rate $f=d\ln D_+(t)/d\ln a$. 

Note that the homogeneous equation of (\ref{eqdel1}) is the same as that of the 
first order one. Therefore, we have the solution of the second order,
\begin{eqnarray}
\delta_2(t,{\bf p})=c_+({\bf p})D_+(t)+c_-({\bf p})D_-(t)+\int_0^t dt'{D_+(t')D_-(t)-D_+(t)D_-(t')\over 
W[D_+(t'),D_-(t')]}S_\delta(t',{\bf p}),
\label{gs}
\end{eqnarray}
where $c_+({\bf p})$ and $c_-({\bf p})$ are constants, and the Wronskian is defined as
\begin{eqnarray}
W[D_+(t),D_-(t)]=D_+(t){\dot D_-(t)}-{\dot D_+(t)}D_-(t).
\end{eqnarray}
In the present paper, we assume the initial density perturbations obey
the Gaussian statistics, and we set $c_\pm({\bf p})=0$. Then, 
the second order solution is written in the form
\begin{eqnarray}
\delta_2(t,{\bf p}) = D_+^2(t) \left(\kappa(t){\cal W}_\alpha({\bf p}) - {2 \over 7}\lambda(t) {\cal W}_\gamma({\bf p})\right),
\label{solutiondelta2}
\end{eqnarray}
with
\begin{eqnarray}
&&\hspace{-1cm}
\kappa(t) 
={1\over D_+^2(t)}\int_0^t{D_-(t)D_+(t')-D_+(t)D_-(t')\over W[D_+(t'),D_-(t')]}D_+^2(t')\left(2f^2H^2-L(t')\right)dt',
\label{kappasolution}
\\
&&\hspace{-1cm}
\lambda(t) 
={7\over 2D_+^2(t)}\int_0^t{D_-(t)D_+(t')-D_+(t)D_-(t')\over W[D_+(t'),D_-(t')]}D_+^2(t')\left(f^2H^2-N_\gamma(t')\right)dt'. 
\label{lambdasolution}
\end{eqnarray}
These expressions are a generalization of the results 
in Ref.\cite{Emilio}.

In the case of the matter dominated universe within the general relativity, 
$a(t)\propto t^{2/3}$, $D_+(t)=a$ and $D_-(t)={a}^{-3/2}$, then
the second order solution reduces to 
\begin{eqnarray}
\delta_2(t,{\bf p}) 
= D_+^2(t)\left({\cal W}_\alpha({\bf p}) - {2\over 7}{\cal W}_\gamma({\bf p})\right).
\end{eqnarray}
Namely, one finds $\kappa(t)=\lambda(t)=1$ in the Einstein de Sitter universe. 
Even in the general second-order scalar-tensor theory, we may consider 
models in which the matter dominated epoch is realized in the early stage of the universe. 
In this stage, the effect of the scalar field perturbations would be negligible, and we 
may naturally expect that the matter density perturbations grow in the same way as those 
in the general relativity. In this case, we may write the initial 
conditions, $\kappa(t)=1$ and $\lambda(t)=1$ at $a\ll1$.

Interestingly, we can show that (\ref{kappasolution}) generally reduces to $\kappa(t)=1$ for 
all the time.
Substituting the expression (\ref{solutiondelta2}) into (\ref{eqdel1}) with regarding 
$\kappa(t)$ and $\lambda(t)$ as unknown functions, we have the following equations
\begin{eqnarray}
&&  \ddot\kappa(t) +(4f+2)H\dot \kappa(t)+(2f^2H^2-L)\kappa(t)=(2f^2H^2-L),
\\
&&  \ddot\lambda(t) +(4f+2)H\dot \lambda(t)+(2f^2H^2-L)\lambda(t)
={7\over 2}(f^2H^2-N_\gamma).
\end{eqnarray}
These equations can be solved, to give the general solutions
\begin{eqnarray}
&&  \kappa(t)=\kappa_+{1\over D_+(t)}+\kappa_-{D_-(t)\over D^2_+(t)}+1,
\\
&&  \lambda(t)=\lambda_+{1\over D_+(t)}+\lambda_-{D_-(t)\over D^2_+(t)}+\lambda_p(t),
\end{eqnarray}
where $\kappa_\pm$ and $\lambda_\pm$ are constants, and $\lambda_p(t)$
is given by the right hand side of (\ref{lambdasolution}).
In the general solutions, we have not imposed the requirement of 
$c_\pm({\bf p})= 0 $ in (\ref{gs}). The condition $c_\pm({\bf p})= 0 $
leads to $\kappa_\pm=\lambda_\pm=0$.
Thus the solutions for $\kappa(t)$ and $\lambda(t)$ are 
$\kappa(t)=1$ and (\ref{lambdasolution}), which we adopt hereafter. 
Therefore, the kernel defined by Eq.~(\ref{kernel}) depends only on the parameter 
$\lambda(t)$, which is determined by the solution of the linear density perturbation, 
$H(t)$ and the function $N_\gamma(t)$. 

Finally, in this section, we present the expression of the velocity divergence at 
the second order of perturbations, which is obtained by 
inserting the expressions of $\delta_1(t,{\bf p})$,
$\theta_1(t,{\bf p})$ and $\delta_2(t,{\bf p})$ into (\ref{cont2}), as
\begin{eqnarray}
\theta_2(t,{\bf p}) &=& D_+^2(t)\left(- \kappa_\theta(t){\cal W}_\alpha({\bf p}) + \lambda_\theta(t){\cal W}_\gamma({\bf p})\right),
\end{eqnarray}
where we defined
\begin{eqnarray}
\kappa_\theta(t) &=& f,\\
\lambda_\theta(t) &=& {4\over 7}f \lambda(t) + {2 \over 7 H}\dot{\lambda}(t).
\end{eqnarray}
In the  Einstein de Sitter universe, we have $\kappa_\theta(t) = \lambda_\theta(t) = 1$.

\section{Bispectrum}
In this section, we consider the bispectrum of the density perturbations
in the most general second-order scalar-tensor theory on the cosmological 
background. The power spectrum and the bispectrum are defined by
\begin{eqnarray}
 &&\left<\delta(t,{\bf k}_1)\delta(t,{\bf k}_2) \right> 
\equiv (2\pi)^3 \delta^{(3)}({\bf k}_1+{\bf k}_2)P(t,k_1),
\\
 &&\left<\delta(t,{\bf k}_1)\delta(t,{\bf k}_2)\delta(t,{\bf k}_3) \right> 
\equiv (2\pi)^3\delta^{(3)}({\bf k}_1+{\bf k}_2+{\bf k}_3)B(t,k_1,k_2,k_3),
\end{eqnarray}
respectively. The three point function at the lowest order of the standard 
perturbation theory is evaluated as
\begin{eqnarray}
\left<\delta(t,{\bf k}_1)\delta(t,{\bf k}_2)\delta(t,{\bf k}_3)\right> 
&&=D^4_+(t)(\left<\delta_{\rm L}({\bf k}_1)\delta_{\rm L}({\bf k}_2)
\delta_{\rm 2K}(t,{\bf k}_3)\right> +2 ~{\rm cyclic~terms}),
\label{ddd} 
\end{eqnarray}
where we defined
\begin{eqnarray}
\delta_{\rm 2K}(t,{\bf k})={\cal W}_\alpha({\bf k}) 
- {2\over 7}\lambda(t){\cal W}_\gamma({\bf k}).
\end{eqnarray}
The first term in the parenthesis in the right hand side of (\ref{ddd}) is
\begin{eqnarray}
&&\hspace{-1cm}\left<\delta_{\rm L}({\bf k}_1)\delta_{\rm L}({\bf k}_2)
\delta_{\rm 2K}(t,{\bf k}_3)\right> 
= \int {d^3q_1\over (2\pi)^3}  F_2(t,{\bf q}_1,{\bf k}_3 - {\bf q}_1) \left<\delta_{\rm L}({\bf k}_1)\delta_{\rm L}({\bf k}_2)
\delta_{\rm L}({\bf q}_1)\delta_{\rm L}({\bf k}_3 - {\bf q}_1)\right>,
\end{eqnarray}
where we defined the kernel
\begin{eqnarray}
F_2(t,{\bf k}_1,{\bf k}_2) \equiv \alpha({\bf k}_1,{\bf k}_2) - {2 \over 7}\lambda(t)\gamma({\bf k}_1,{\bf k}_2). 
\label{kernel}
\end{eqnarray}
Using the definition of the linear matter power spectrum, 
\begin{eqnarray}
 \left<\delta_{\rm L}({\bf k}_1)\delta_{\rm L}({\bf k}_2) \right> = (2\pi)^3 \delta^{(3)}({\bf k}_1+{\bf k}_2)P_{11}(k_1),
\end{eqnarray}
and the Wick's theorem, 
we have
\begin{eqnarray}
&&\left<\delta_{\rm L}({\bf k}_1)\delta_{\rm L}({\bf k}_2)\delta_{\rm 2K}(t,{\bf k}_3)\right> 
= 2(2\pi)^3  \delta^{(3)}({\bf k}_1+{\bf k}_2 + {\bf k}_3 )
F_2(t,{\bf k}_1,{\bf k}_2) P_{11}(k_1) P_{11}(k_2),
\end{eqnarray}
where we may use
$\alpha({\bf k}_1,{\bf k}_2)=\alpha(-{\bf k}_1,-{\bf k}_2)$, 
$\gamma({\bf k}_1,{\bf k}_2)=\gamma(-{\bf k}_1,-{\bf k}_2)$,
$F_2(t,{\bf k}_1,{\bf k}_2) = F_2(t,-{\bf k}_1,-{\bf k}_2)= 
F_2(t,{\bf k}_2,{\bf k}_1)$. 
Finally, we have the expression for the bispectrum at the lowest order of the 
perturbation theory, 
\begin{eqnarray}
B(t,k_1,k_2,k_3)=D_+^4(t) B_4(t,k_1,k_2,k_3)
\end{eqnarray}
with 
\begin{eqnarray}
&&B_4 (t,k_1,k_2,k_3) = 2F_2(t,{\bf k}_1,{\bf k}_2)P_{11}(k_1) P_{11}(k_2)
+2{\rm ~cyclic~terms}.
\end{eqnarray}
The reduced bispectrum is given by

\begin{figure}
 \begin{minipage}{0.5\hsize}
  \begin{center}
  \vspace{-5.8cm}
  \includegraphics[width=130mm,bb=0 0 640 480]{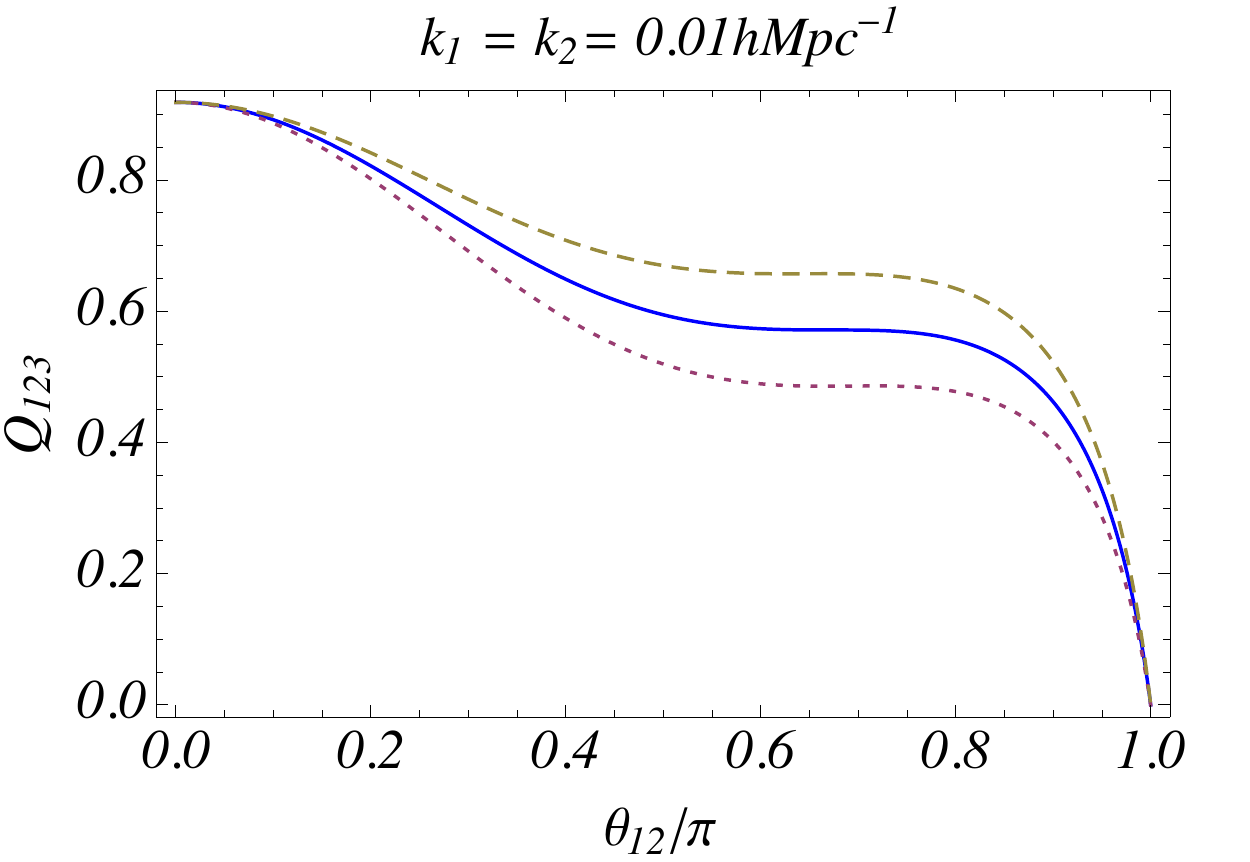}
  \end{center}
   \vspace{-0.8cm}
  \label{fig:one}
 \end{minipage}\begin{minipage}{0.5\hsize}
  \begin{center}
    \vspace{-5.8cm}
   \includegraphics[width=130mm,bb=0 0 640 480]{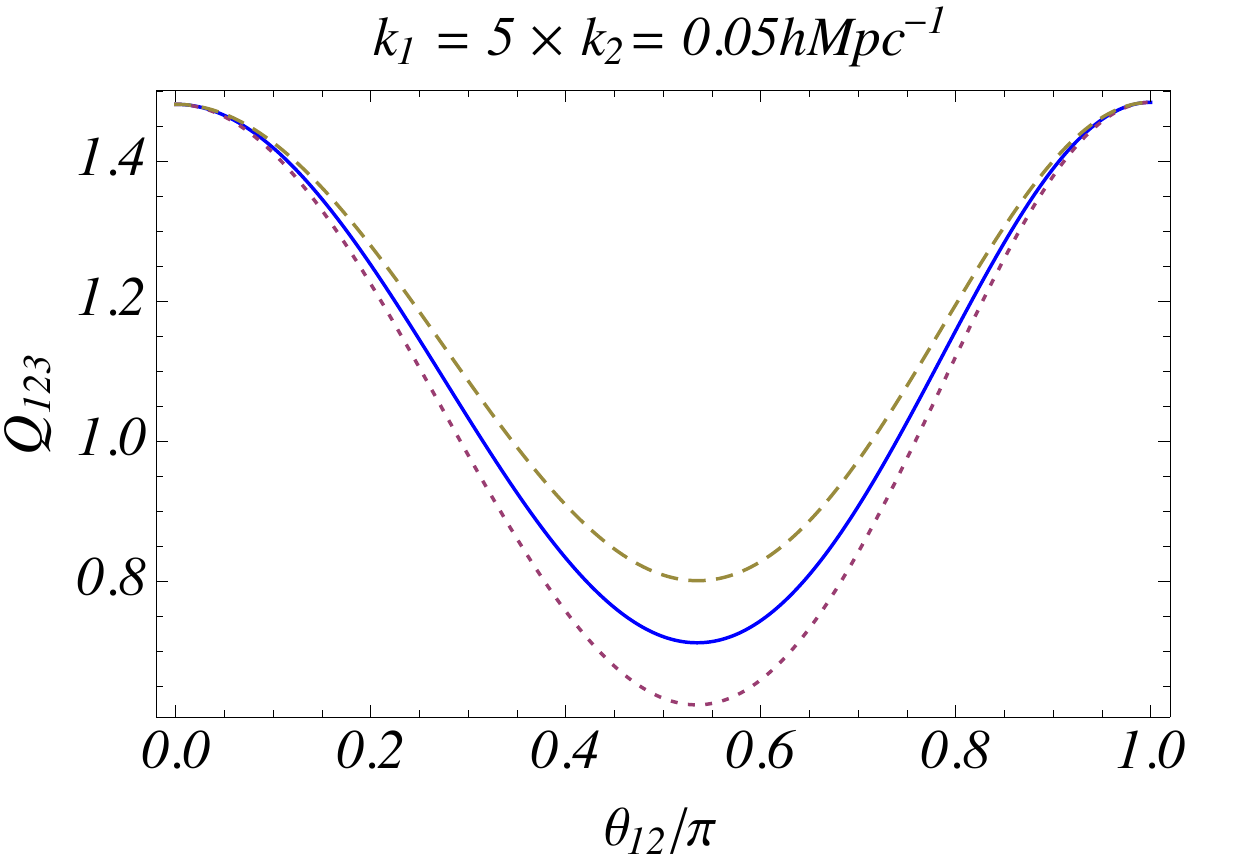}
  \end{center}
   \vspace{-0.8cm}
  \label{fig:two}
 \end{minipage}
 \begin{minipage}{0.5\hsize}
  \begin{center}
  \vspace{-4cm}
  \includegraphics[width=130mm,bb=0 0 640 480]{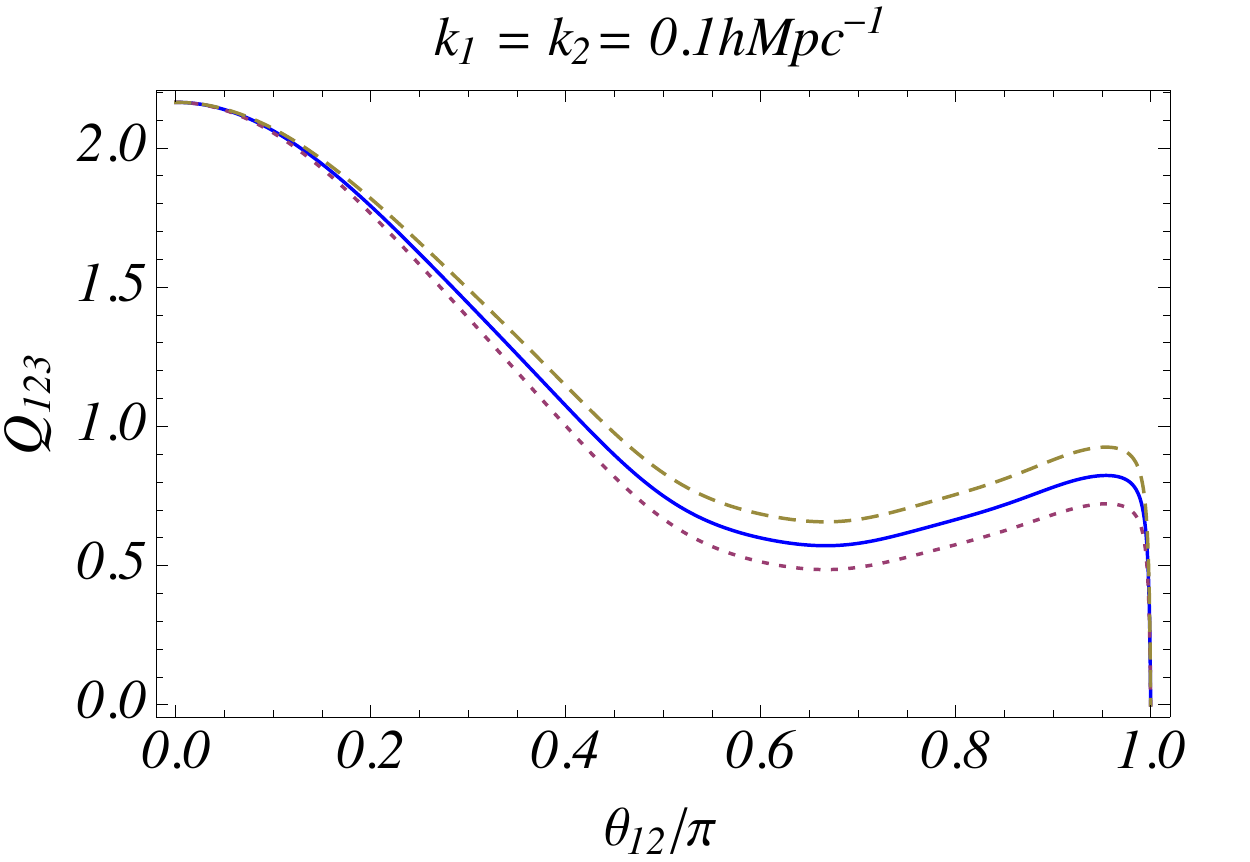}
  \end{center}
   \vspace{-0.8cm}
  \label{fig:three}
 \end{minipage}\begin{minipage}{0.5\hsize}
  \begin{center}
    \vspace{-4cm}
   \includegraphics[width=130mm,bb=0 0 640 480]{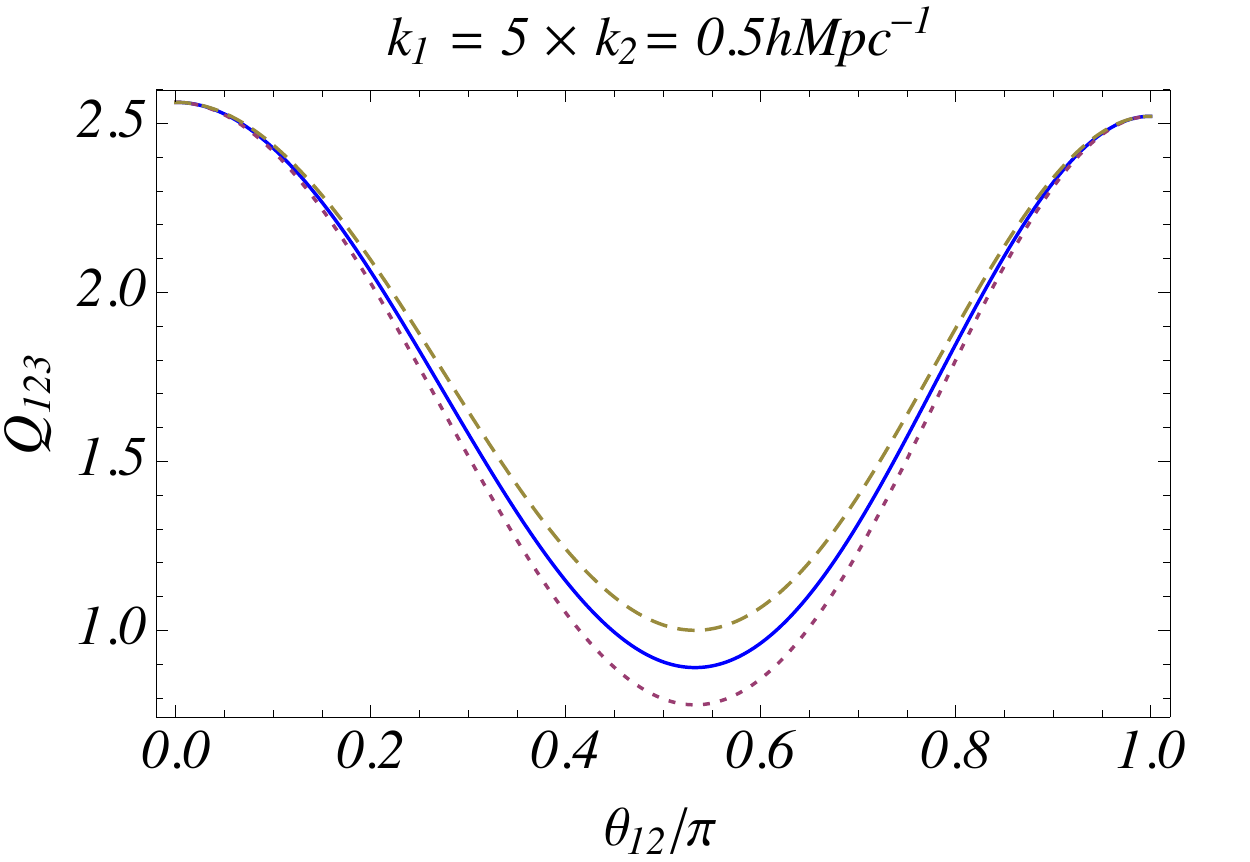}
  \end{center}
   \vspace{-0.0cm}
\label{fig:four}
 \end{minipage}
 \caption{$Q_{123}$ as function of $\theta_{12}$ with fixing,
$k_1=k_2=0.01h{\rm Mpc}^{-1}$ (upper left panel),
$k_1=k_2=0.1h{\rm Mpc}^{-1}$ (lower left panel),
$k_1=5\times k_2=0.05h{\rm Mpc}^{-1}$ (lower right panel),
and
$k_1=5\times k_2=0.5h{\rm Mpc}^{-1}$ (lower right panel), respectively. 
For the linear matter power spectrum $P_{11}(k)$, 
we adopt the spatially flat universe with the cold dark matter model (CDM) and 
the cosmological constant $\Lambda$, whose density parameters are $\Omega_0=0.3$ and 
$\Omega_\Lambda=0.7$, respectively. 
Note that the reduced bispectrum depends on time $t$ only through $\lambda(t)$, 
for which we adopted the different value of $\lambda(t)=1$ (blue solid curve), 
$\lambda(t)=1.2$ (red dotted curve), and $\lambda(t)=0.8$ (yellow dashed curve), 
irrespectively of the $\Lambda$CDM model. 
}
\end{figure}

\begin{eqnarray}
Q_{123}(t,k_1,k_2,\theta_{12}) &=&{B_4(t,k_1,k_2,k_3)\over P_{11}(k_1)P_{11}(k_2) + 
P_{11}(k_2)P_{11}(k_3) + P_{11}(k_3)P_{11}(k_1)},
\end{eqnarray}
at the lowest order of perturbations. Note that the (reduced) bispectrum 
is described by the kernel (\ref{kernel}), which depends only on the parameter 
$\lambda(t)$, given by (\ref{lambdasolution}).

Because ${\bf k}_1+{\bf k}_2+{\bf k}_3=0$ is satisfied, and the reduced bispectrum 
is a function of only three parameters, which we take $k_1=|{\bf k}_1|$, 
$k_2=|{\bf k}_2|$ and the angle $\theta_{12}$ between  ${\bf k}_1$ and ${\bf k}_2$.  
Explicit expressions for $\alpha({\bf k}_i,{\bf k}_j)$ and 
$\gamma({\bf k}_i,{\bf k}_j)$, where $(i,j)$ denotes any of 
$(1,2),~(2,3)$, or $(3,1)$, are summarized in Appendix B. 

Each panel of figure 1 shows a typical behavior of $Q_{123}$ as function of $\theta_{12}$ with fixing 
$k_1$ and $k_2$, whose values are described in the caption.
In each panel, we adopt the different value of $\lambda(t)=1$ (blue solid curve), 
$\lambda(t)=1.2$ (red dotted curve), and $\lambda(t)=0.8$ (yellow dashed curve), where 
we assumed the spatially flat universe with the cold dark matter model (CDM) and the cosmological 
constant $\Lambda$, whose density parameters are $\Omega_0=0.3$ and $\Omega_\Lambda=0.7$,
for the linear matter power spectrum $P_{11}(k)$. 
Note that the reduced bispectrum depends on time $t$ only through $\lambda$(t). 
One can read the following features. First, 
the overall amplitude of $Q_{123}$ depends on the value of $k_1$ and $k_2$. 
However, once the values of $k_1$ and $k_2$ are fixed, the reduced bispectrum 
is enhanced for $\lambda<1$, while it is reduced for $\lambda>1$. 
This feature is explained by the expression of kernel (\ref{kernel})
and the fact $\gamma({\bf k}_i,{\bf k}_j)\geq0$.

In the limit $\theta_{12} = 0$, we have $\gamma({\bf k}_1,{\bf k}_2)=\gamma({\bf k}_2,{\bf k}_3)=\gamma({\bf k}_3,{\bf k}_1)=0$ (see also appendix B). Then, $Q_{123}$ is independent of $\lambda$ at 
$\theta_{12} = 0$.
In the limit $\theta_{12} = \pi$, 
$Q_{123}$ has the different behavior depending on the conditions $k_1=k_2$ and $k_1\neq k_2$.
In the case $k_1 \neq k_2$, we have 
$\gamma({\bf k}_1,{\bf k}_2)=\gamma({\bf k}_2,{\bf k}_3)=\gamma({\bf k}_3,{\bf k}_1)=0$, 
which is the same as those  of the limit $\theta_{12} = 0$. In the case $k_1=k_2$, however, we have 
$\gamma({\bf k}_1,{\bf k}_2)=0,~\gamma({\bf k}_2,{\bf k}_3)=\gamma({\bf k}_3,{\bf k}_1)=1$ 
, and $k_3=0$, i.e, $P_{11}(k_3) = 0$. Then the bispectrum approach zero in this limit, 
thought the rate of convergence depends on $\lambda(t)$, as is 
discussed in the next section.

All the influence of the nonlinear interaction of the modified gravity arise only through 
the parameter $\lambda(t)$, which appears as the term in proportion to $\gamma({\bf k}_1,{\bf k}_2)$
in the kernel (\ref{kernel}). 
The bispectrum of the matter density perturbations behaves only in a restricted way, which is 
a feature of the general second-order scalar-tensor theory equipped with the Vainshtein mechanism.

\section{Kinetic gravity braiding model}
In this section, we consider a simple example to demonstrate how the modification 
of gravity influences the behavior of the bispectrum at a quantitative level. 
We consider the kinetic gravity braiding model investigated in Ref.~\cite{Deffayet,kgb}, 
whose action is written as 
\begin{eqnarray}
S=\int d^4 x\sqrt{-g}\left[{M_{\rm pl}^2\over 2}R+K-G_3\square \phi+{\cal L}_{\rm m}\right], 
\label{kgbaction}\end{eqnarray}
with the Planck mass $M_{\rm pl}$, which is related with the gravitational constant $G_N$
by $8\pi G_N=1/M_{\rm pl}^2$. Comparing this action (\ref{kgbaction}) with that of the 
most general second-order scalar-tensor theory, the action of the kinetic gravity 
braiding model is produced by setting 
\begin{eqnarray}
&&G_4={M_{\rm pl}^2\over 2},~~~~G_5=0. 
\end{eqnarray}
In Ref.~\cite{kgb}, $K$ and $G_3$ are chosen as
\begin{eqnarray}
&&K=-X, ~~~~G_3=M_{\rm pl}\left({r_c^2\over M_{\rm pl}^2}X\right)^n,  
\end{eqnarray}
where $n$ and $r_c$ are the parameters. 
In this model, we have 
\begin{eqnarray}
&&L(t)=-{A_0 {\cal F}_T \rho_{\rm m}\over 2(A_0 {\cal G}_T + A_2^2 {\cal F}_T)},\\
&&N_{\gamma}(t)={B_0 A_2^3 {\cal F}_T^3 \rho_{\rm m}^2 \over 4(A_0 {\cal G}_T^2 + A_2^2 {\cal F}_T)^3H^2}.
\end{eqnarray}
Useful expressions of the kinetic gravity braiding model are summarized in Appendix A.
\begin{figure}[t]
\begin{flushright}
  \includegraphics[width=120mm,bb=0 0 500 240]{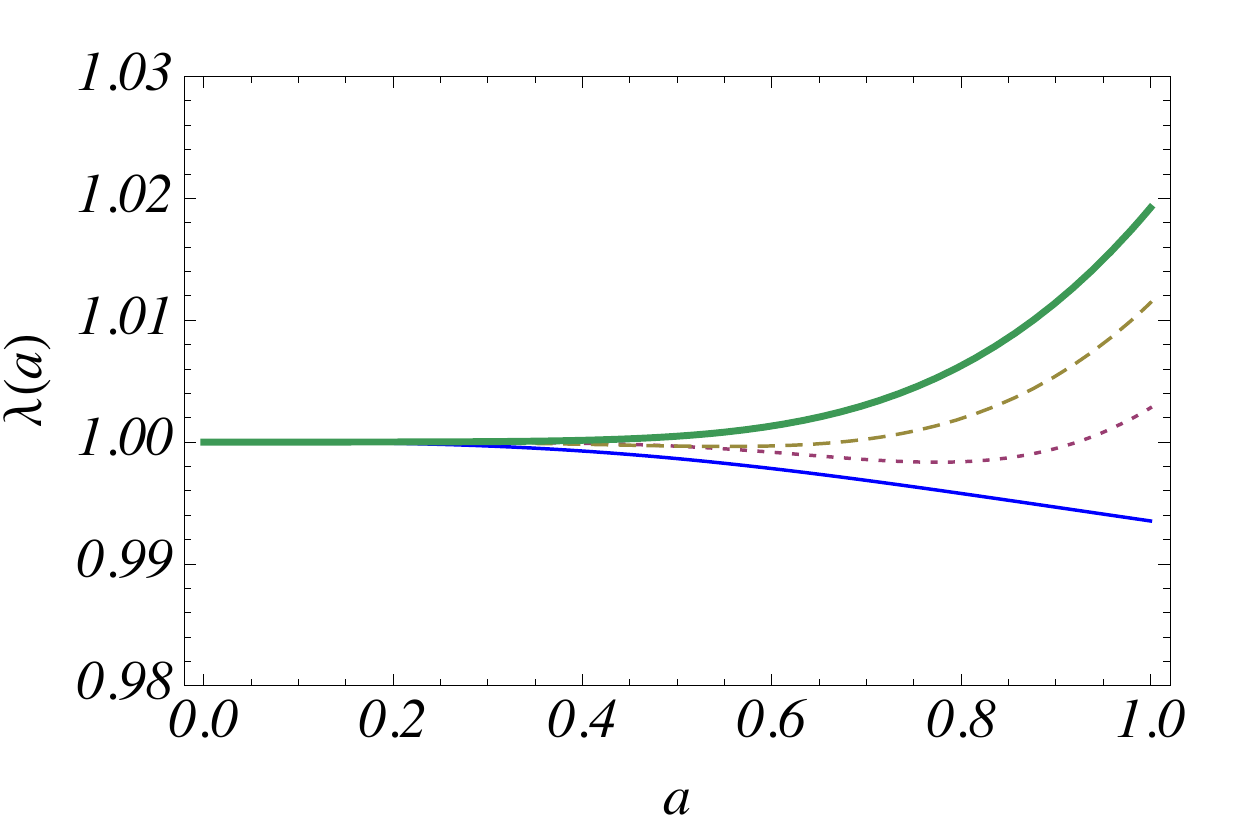}
  \caption{
$\lambda(t)$ as function of $a$ for the $\Lambda$CDM model 
(blue solid curve) and the kinetic gravity braiding model with 
$n=1$ (red dotted curve), $n=2$ (yellow dashed curve), and $5$ 
(green thick curve).}
\end{flushright}
\end{figure}

When we consider the attractor solution, 
which satisfies 
\begin{eqnarray}
3\dot\phi H G_{3X}=1, 
\label{attractor}
\end{eqnarray}
the Friedmann equation is written in the form 
\begin{eqnarray}
\left({H\over H_0}\right)^2={\Omega_0\over a^3}
+(1-\Omega_0)\left({H\over H_0}\right)^{-2/(2n-1)},
\end{eqnarray}
where $H_0$ is the Hubble constant and 
$\Omega_0$ is the density parameter
at the present time, and the model parameters must satisfy
\begin{eqnarray}
H_0r_c=\left({2^{n-1}\over 3n}\right)^{1/2n}\left[{1\over 6(1-\Omega_0)}\right]^{(2n-1)/4n}. 
\end{eqnarray}
On the attractor solution, $L(t)$ and $N_\gamma(t)$ reduce to
\begin{eqnarray}
&&L(t)=-{3\over 2}{2n + (3n -1)\Omega_m(t)\over 5n-\Omega_m(t)}H^2,\\
&&N_{\gamma}(t)=-{9\over 4}{(1 - \Omega_m(t))(2n - \Omega_m(t))^3\over \Omega_m(t)(5n - 
\Omega_m(t))^3}H^2,
\end{eqnarray}
where $\Omega_m(a)$ is defined by $\Omega_m(a)={\Omega_0H_0^2/H(a)^2a^3}$. 
Note that the quasi-static approximation on the scales of the large scale 
structures  holds for $n\simlt 10$ (see \cite{kgb}).

Figure 2 shows the evolution of $\lambda(t)$ as a function of $a$ 
for the kinetic gravity braiding model with 
$n=1,~2,~5$ and the $\Lambda$CDM model. 
For $a\ll1$, we have $\lambda(t)=1$, which is the prediction of the 
Einstein de Sitter universe.
However, the accelerated expansion arises due to a domination of  
the galileon field as $a$ approaches $1$, then the value of 
$\lambda(t)$ starts to deviate from $1$. 
The deviation of $\lambda(t)$ from $1$ is small. 
The value of $\lambda(t)$ at the present epoch is $0.994$
for the $\Lambda$CDM model with the density parameter $\Omega_0=0.3$. 
The value of $\lambda(t)$
at the present epoch is $1.003$, $1.011$, and $1.019$ for the KBG model 
with $n=1,~2,~5$, respectively. 
Our results guarantee the validity of the approximation setting 
$\lambda(t)=1$, which is usually adopted in the standard density perturbations theory.

Figure 3 shows the relative deviation of the bispectrum at the present epoch 
of the KGB model 
from that of the $\Lambda$CDM model, $Q_{123}(t,k_1,k_2,\theta_{12})/$ $Q_{123\Lambda}(t,k_1,k_2,\theta_{12})-1$, 
as a function of $\theta_{12}$, where 
$Q_{123\Lambda}(t,k_1,k_2,\theta_{12})$ is the reduced bispectrum of the $\Lambda$CDM model.
The relative deviation from the $\Lambda$CDM model is less than $2$~\%.
For the case $k_1\neq k_2$, the deviation between the models does not appear 
at $\theta_{12}=0,~\pi$, which is simply understood by the fact 
$\gamma({\bf k}_i,{\bf k}_j)=0$ there.  
In the case $k_1=k_2$ in the limit $\theta_{12} = \pi$, we have 
$\alpha({\bf k}_1,{\bf k}_2)\sim(\pi-\theta_{12})^2$, 
$\alpha({\bf k}_2,{\bf k}_3)=\alpha({\bf k}_3,{\bf k}_1)=3/4$, 
$\gamma({\bf k}_1,{\bf k}_2)\sim(\pi-\theta_{12})^2$, 
$\gamma({\bf k}_2,{\bf k}_3)=\gamma({\bf k}_3,{\bf k}_1)=1$, 
and 
$P(k_3)\propto k_3^{n_s}\propto (\pi-\theta_{12})^{n_s}$, where $n_s$ is the spectral index.
(see appendix B for details.)
\begin{figure}[t!]
 \begin{minipage}{0.5\hsize}
  \begin{center}
  \vspace{-5.8cm}
  \includegraphics[width=130mm,bb=0 0 640 480]{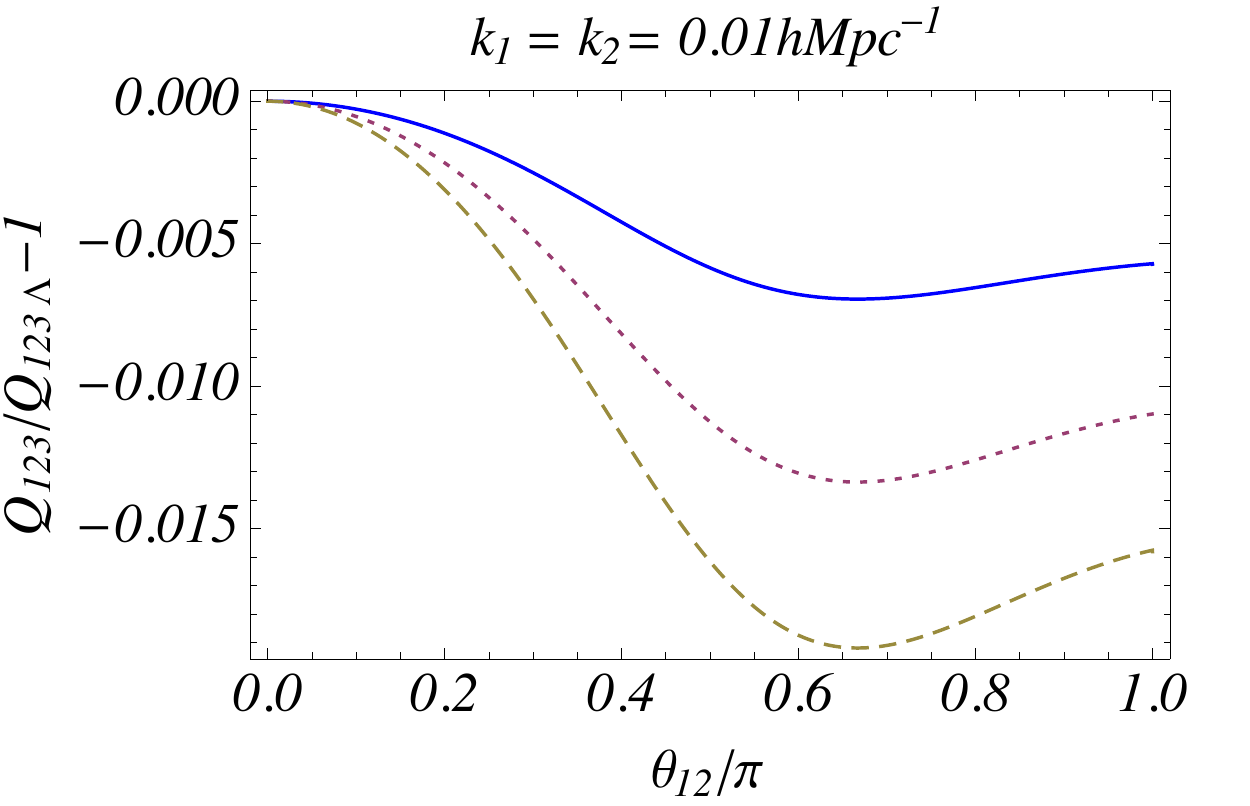}
  \end{center}
   \vspace{-0.8cm}
  \label{fig:one}
 \end{minipage}\begin{minipage}{0.5\hsize}
  \begin{center}
    \vspace{-5.8cm}
   \includegraphics[width=130mm,bb=0 0 640 480]{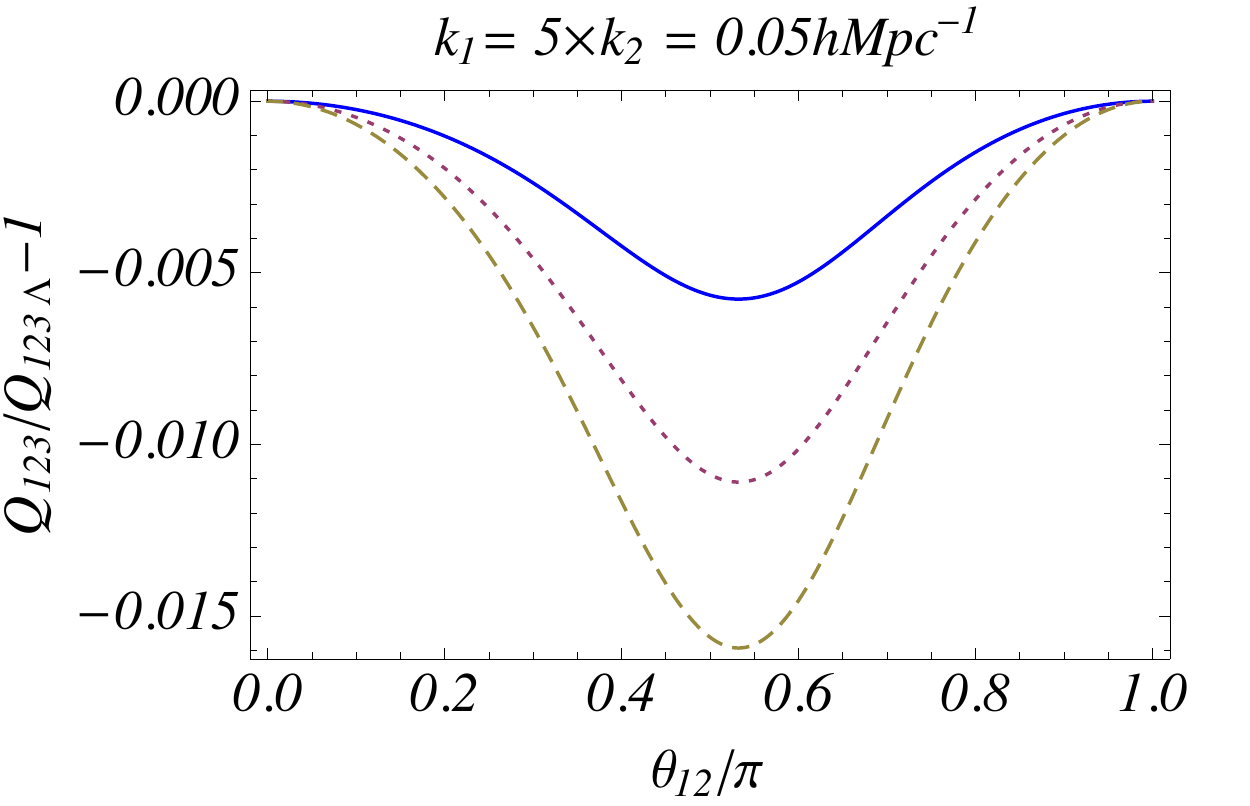}
  \end{center}
   \vspace{-0.8cm}
  \label{fig:two}
 \end{minipage}
 \begin{minipage}{0.5\hsize}
  \begin{center}
  \vspace{-4cm}
  \includegraphics[width=130mm,bb=0 0 640 480]{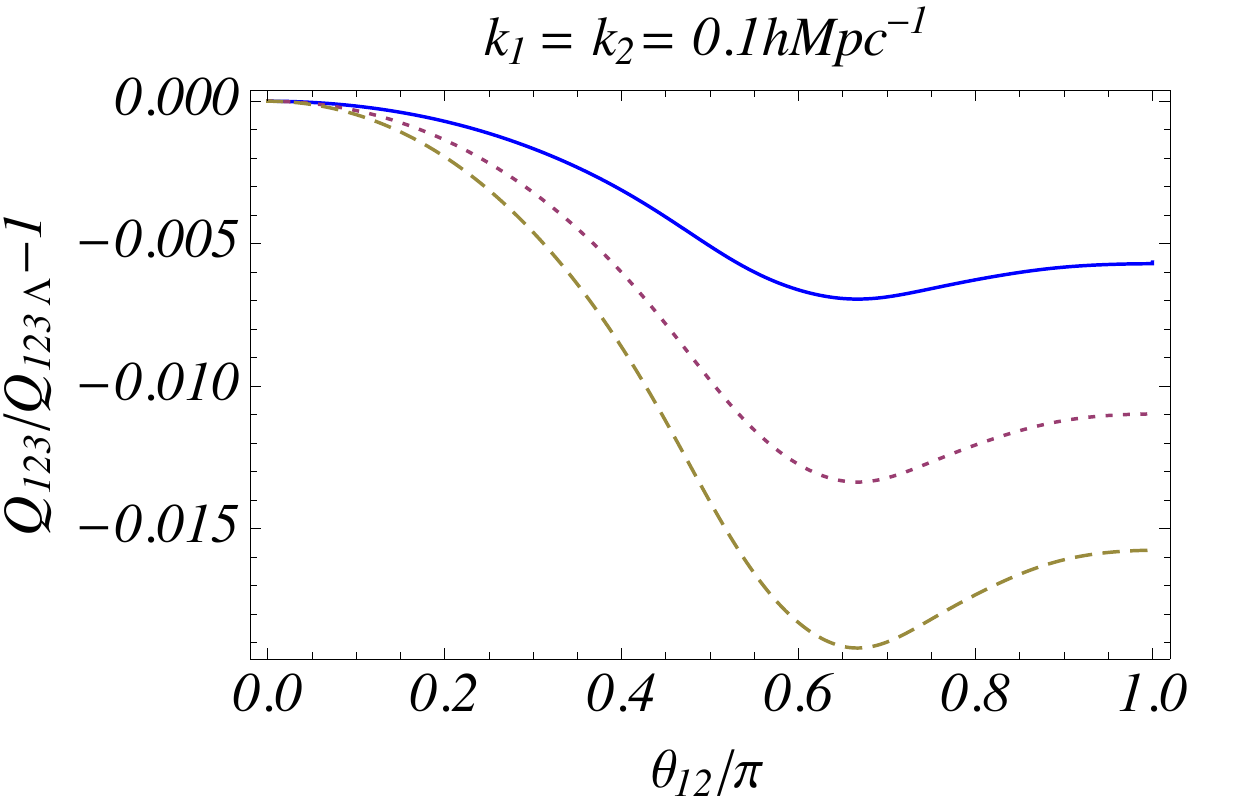}
  \end{center}
   \vspace{-0.3cm}
  \label{fig:three}
 \end{minipage}\begin{minipage}{0.5\hsize}
  \begin{center}
    \vspace{-4cm}
   \includegraphics[width=130mm,bb=0 0 640 480]{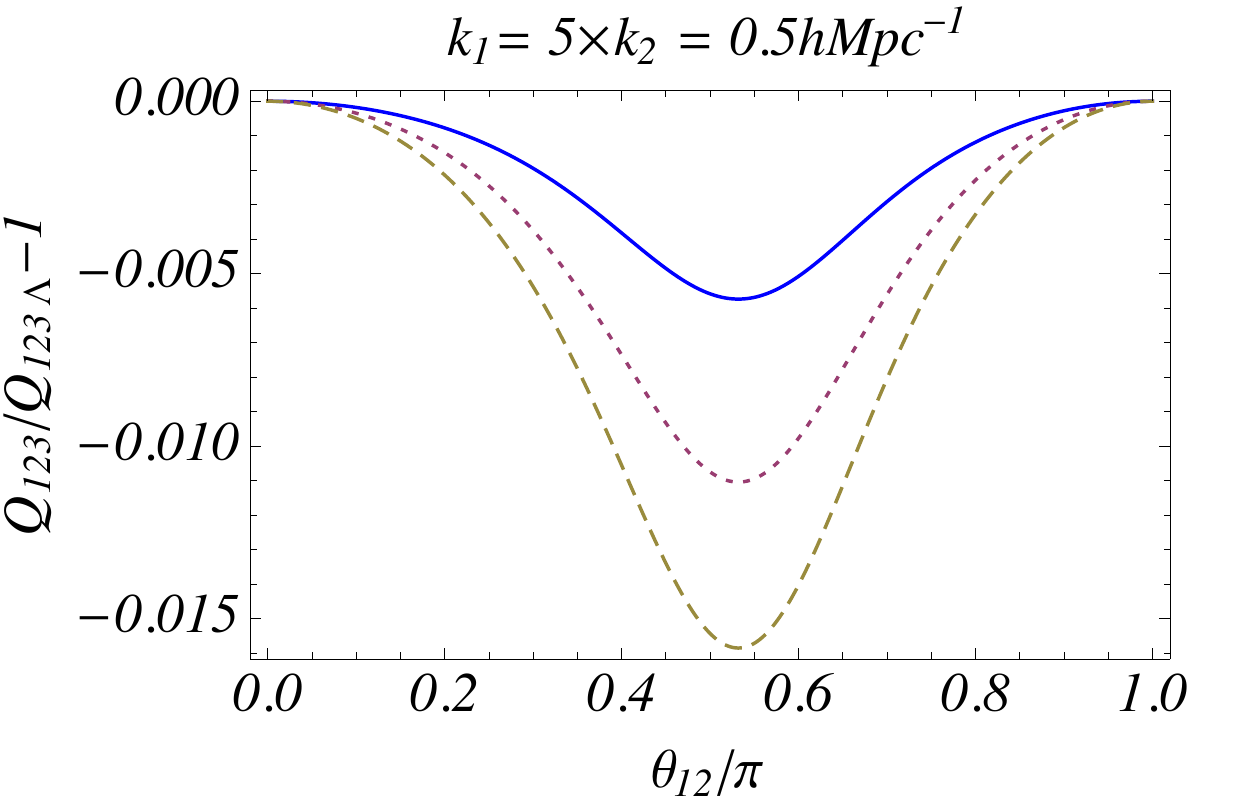}
  \end{center}
   \vspace{-0.3cm}
\label{fig:four}
 \end{minipage}
 \caption{Relative deviation of the reduced bispectrum at the present epoch 
of the kinetic
gravity braiding model with $n=1$ (blue solid curve), $n=2$ (red dotted curve),
$n=5$ (yellow dashed curve) from the that of 
the $\Lambda$CDM model $Q_{123\Lambda}$, as a function 
of $\theta_{12}$, where $k_1$ and $k_2$ are fixed, whose values are noted 
on each panel. Here the density parameter is fixed as $\Omega_0=0.3$.}
\end{figure}
Then, the bispectrum has the asymptotic form
\begin{eqnarray}
B_4(t,k_1,k_1,\theta_{12})
\sim4\left({3\over 4} - {2\over 7}\lambda(t)\right)P_{11}(k_3)P_{11}(k_1)
\end{eqnarray}
around the limit $\theta_{12} = \pi$. This leads to the ratio of the 
reduced bispectrum in this limit,
\begin{eqnarray}
{Q_{123}(t,k_1,k_1,\theta_{12})\over Q_{123\Lambda}(t,k_1,k_1,\theta_{12})}
={21-8\lambda(t)\over 21-8\lambda_\Lambda(t)},
\end{eqnarray}
where $\lambda_\Lambda(t)$
is the parameter $\lambda(t)$ of the $\Lambda$CDM model,
which explains the behaviors of the left panels of Fig.~3.

The behavior of the reduced bispectrum is almost same when the ratio of 
$k_1/k_2$ is the same. This is because the function 
$\alpha({\bf k}_i,{\bf k}_j)$ and $\gamma({\bf k}_i,{\bf k}_j)$ 
depend on only on the ratio $k_1/k_2$ and $\theta_{12}$ (see also appendix B). 

Recently, the bispectrum in the covariant cubic galileon cosmology is 
investigated in Ref.~\cite{Emilio}. 
Our kinetic gravity braiding model with $n=1$ is a cubic galileon model, 
however, there is the difference between our model and the covariant 
cubic galileon cosmology in Ref.~\cite{Emilio}. 
The cosmic acceleration in the covariant cubic galileon model 
is derived by a potential of the scalar field. 
This causes the differences of the evolution of the background universe
and the linear density perturbations.

\section{Summary and Conclusions}
In the present paper, we have investigated the bispectrum of the matter density 
perturbations induced by the gravitational instability in the most general 
second-order scalar-tensor theory that may possess the Vainshtein mechanism. 
We have discussed a general feature of this wide class of modified gravity models in
the most general second-order scalar-tensor theory.
We have obtained the expression of the bispectrum of the 
second order of perturbations on the basis of the standard 
density perturbation theory in an analytic manner. 
The bispectrum is expressed by the kernel (\ref{kernel}), depending only 
on the parameter $\lambda(t)$, which is determined by the growing and decaying 
solutions of the linear density perturbations $D_\pm(t)$, the Hubble parameter $H(t)$, 
and the other function $N_\gamma(t)$ for the nonlinear interactions.
These simple results come from the fact that the basic equations for the 
gravitational and scalar fields have the same form of the nonlinear mode 
couplings, which are derived as the leading terms under the quasi-static 
approximation within the subhorizon scales.
Thus, all the effect of the modified gravity in the bispectrum come through 
the parameter $\lambda(t)$ in the kernel (\ref{kernel}), which has the simple 
structure. This makes the behavior of the bispectrum less complex. 
As an application of our results, we have exemplified the behavior of the 
bispectrum in a kinetic gravity braiding model proposed in Ref.~\cite{kgb}.
We have investigated the evolution of $\lambda(t)$ in this model, and 
have demonstrated the deviation of the reduced bispectrum from that of the 
$\Lambda$CDM model is less than a few \%. 
Higher order solutions of the density perturbations will be obtained in a 
similar way, which is left as a future problem.

\section*{Acknowledgment}
We thank R. Kimura, T. Kobayashi, and A. Taruya for useful discussions 
at the early stage of this work. 
K.Y. thanks Prof. L. Amendola for the hospitality during his stay at Heidelberg University. 
He also thanks E. Bellini and M. Takada for useful communications related 
to the topic of the present paper. 
This work is supported by exchange visits between JSPS and DFG.
The research by K.Y. is supported in part by Grant-in-Aid for Scientific researcher
of Japanese Ministry of Education, Culture, Sports, Science and Technology 
(No.~21540270 and No.~21244033).

\appendix 
\section{Definition of the coefficients}
We first summarize the definitions of the coefficients in the field equations
in section 2. 
\begin{eqnarray}
A_0&:=&\frac{\dot\Theta}{H^2}+\frac{\Theta}{H}
+{\cal F}_T-2{\cal G}_T-2\frac{\dot{\cal G}_T}{H}-\frac{{\cal E}+{\cal P}}{2H^2},
\\
A_1&:=&\frac{\dot{\cal G}_T}{H}+ {\cal G}_T-{\cal F}_T,
\\
A_2&:=& {\cal G}_T-\frac{\Theta}{H},
\\
B_0&:=&\frac{X}{H}\biggl\{\dot\phi G_{3X}+3\left(\dot X+2HX\right)G_{4XX}
+2X\dot XG_{4XXX}-3\dot\phi G_{4\phi X}
+2\dot\phi XG_{4\phi XX}
\nonumber\\&&
+\left(\dot H+H^2\right)\dot\phi G_{5X}
+\dot\phi
\left[2H\dot X+\left(\dot H+H^2\right) X\right]G_{5XX}
+H\dot\phi X\dot XG_{5XXX}
\nonumber\\&&
-2\left(\dot X+2HX\right)G_{5\phi X}
-\dot\phi XG_{5\phi\phi X}-X\left(\dot X-2HX\right)G_{5\phi XX}\biggr\},
\\
B_1&:=&2X\left[G_{4X}+\ddot\phi\left(G_{5X}+XG_{5XX}\right)
-G_{5\phi}+XG_{5\phi X}\right],
\\
B_2&:=&
-2X\left(G_{4X}+2XG_{4XX}+H\dot\phi G_{5X}
+H\dot\phi XG_{5XX}-G_{5\phi}-XG_{5\phi X}\right),
\\
B_3&:=&H\dot\phi XG_{5X},
\\
C_0&:=&2X^2G_{4XX}+\frac{2X^2}{3}\left(2\ddot\phi G_{5XX}
+\ddot\phi XG_{5XXX}-2G_{5\phi X}+XG_{5\phi XX}\right),
\\
C_1&:=&H\dot\phi X\left(G_{5X}+XG_{5XX}\right),
\end{eqnarray}
where we also defined
\begin{eqnarray}
{\cal F}_T&:=&2\left[G_4
-X\left( \ddot\phi G_{5X}+G_{5\phi}\right)\right],
\\
{\cal G}_T&:=&2\left[G_4-2 XG_{4X}
-X\left(H\dot\phi G_{5X} -G_{5\phi}\right)\right],
\\
\Theta&:=&-\dot\phi XG_{3X}+
2HG_4-8HXG_{4X}
-8HX^2G_{4XX}+\dot\phi G_{4\phi}+2X\dot\phi G_{4\phi X}
\nonumber\\&&
-H^2\dot\phi\left(5XG_{5X}+2X^2G_{5XX}\right)
+2HX\left(3G_{5\phi}+2XG_{5\phi X}\right),
\\
{\cal E} &:=& 2 X K_X - K + 6 X \dot \phi  H G_{3X} - 2 X G_{3 \phi} - 6 H^2 G_4
 + 24 H^2 X (G_{4X} + X G_{4XX})
\nonumber\\&&
 - 12 H X \dot \phi G_{4 \phi X}- 6 H \dot \phi  G_{4\phi} + 2 H^3 X \dot \phi (5 G_{5 X} 
 + 2 X G_{5XX}) 
\nonumber\\&&
- 6 H^2 X (3 G_{5 \phi} + 2 X G_{5\phi X}),
\\
{\cal P}&:=& K - 2X(G_{3 \phi} + \ddot \phi G_{3 X}) + 2(3 H^2 + 2 \dot H)G_4 
- 12 H^2 X G_{4X} - 4 H \dot X G_{4 X} 
\nonumber\\&&
- 8 \dot H X G_{4X} - 8 H X \dot X G_{4 X X} 
+ 2 (\ddot \phi + 2 H \dot \phi)G_{4 \phi} + 4 X G_{4 \phi \phi} + 4 X (\ddot \phi
 - 2 H \dot \phi)G_{4\phi X} 
\nonumber\\&&
- 2 X (2 H^3 \dot \phi + 2 H \dot H \dot \phi + 3 H^2 \ddot \phi)G_{5 X}
- 4 H^2 X^2 \ddot \phi G_{5XX} + 4 H X (\dot X - H X)G_{5 \phi X} 
\nonumber\\&&+ 2\left[ 2 (H X) \dot{}
 + 3 H^2 X \right]G_{5\phi} + 4 H X \dot \phi G_{5 \phi \phi}.
\end{eqnarray}
In the kinetic gravity braiding model considered in section 4, 
the coefficients are written as follows,
\begin{eqnarray}
&&{\cal F}_T = M_{pl}^2,~~~~~{\cal G}_T =M_{pl}^2,\\
&&\Theta = - n M_{pl}\left({r_c^2 \over M_{pl}^2}\right)^n \dot{\phi} X^n + H M_{pl}^2,\\
&&\dot{\Theta} = -n(2n + 1)M_{pl}\left({r_c^2 \over M_{pl}^2}\right)^n \ddot{\phi} X^n + \dot{H}M_{pl}^2,\\
&&{\cal E}= - X + 6 n M_{pl}\left({r_c^2 \over M_{pl}^2}\right)^n \dot{\phi}H X^n - 3 H^2 M_{pl}^2,\\
&&{\cal P}= - X - 2n M_{pl}\left({r_c^2 \over M_{pl}^2}\right)^n\ddot{\phi}X^n + (3 H^2 + 2\dot{H})M_{pl}^2,\\
&&A_0 = {X\over H^2} - 2nM_{pl}\left({r_c^2 \over M_{pl}^2}\right)^n \left({2\dot{\phi}\over H} + n{\ddot {\phi}\over H^2}\right)X^n,\\
&&A_2 = B_0 = n {\dot{\phi}\over H}M_{pl}\left({r_c^2 \over M_{pl}^2}\right)^n X^n,\\
&&A_1 = B_1 = B_2 = B_3 = C_0 = C_1 = 0.
\end{eqnarray}
In the present papwer, we consider the attractor solution 
satisfying (\ref{attractor}), 
then we have
\begin{eqnarray}
&&\ddot{\phi} = - {1 \over 2n-1}{\dot{\phi} \dot{H}\over H},\\
&&{\dot{H} \over H^2} = - {(2n-1) 3 \Omega_m(a) \over 2(2n - \Omega_m(a))},
\\
&&A_0 
=-{M_{\rm Pl}^2(1-\Omega_m(a))\left(2n +(3n - 1)\Omega_m(a)\right) 
\over 2n-\Omega_m(a)},\\
&&A_2=
M_{\rm Pl}^2{(1-\Omega_m(a))},\\
&&B_0=
M_{\rm Pl}^2{(1-\Omega_m(a))},
\end{eqnarray}
where we defined $\Omega_m(a)=\rho_m(a)/3M_{\rm Pl}^2H^2$.

\section{Explicit expressions of $\alpha$ and $\gamma$}
In general, we may write the wave number vector, which satisfies
${\bf k}_1 + {\bf k}_2 + {\bf k}_3 = 0$, as follows,
\begin{eqnarray}
{\bf k}_1 &=& (0,0,k_1),\\
{\bf k}_2 &=& (0, k_2 \sin \theta_{12},k_2 \cos \theta_{12}),\\
{\bf k}_3 &=& (0,-k_2 \sin \theta_{12},-k_1 - k_2 \cos \theta_{12}),
\end{eqnarray}
where $\theta_{12}$ is the angle between the vector ${\bf k}_1$ 
and ${\bf k}_2$. Then we have
\begin{eqnarray}
&&{{\bf k}_1 \cdot {\bf k}_2 \over k_1 k_2}=\cos\theta_{12},
\\
&&{{\bf k}_2 \cdot {\bf k}_3 \over k_2 k_3}
 = {-k_2 - k_1 \cos \theta_{12}\over \sqrt{k_1^2 + k_2^2 + 2 k_1 k_2 \cos \theta_{12}}},\\
&&{{\bf k}_3 \cdot {\bf k}_1 \over k_3 k_1}
 = {-k_1 - k_2 \cos \theta_{12}\over \sqrt{k_1^2 + k_2^2 + 2 k_1 k_2 \cos \theta_{12}}},
\end{eqnarray}
where we used  $k_3 = \sqrt{k_1^2 + k_2^2 + 2 k_1 k_2 \cos \theta_{12}}$. 
Introducing the constant $c$ by $k_1 = c k_2$, we have
\begin{eqnarray}
&&k_3
=k_1\sqrt{c^2 + 2c \cos \theta_{12} + 1},
\\
&&{{\bf k}_2 \cdot {\bf k}_3 \over k_2 k_3}
=-{c +  \cos \theta_{12} \over \sqrt{c^2 + 2c \cos \theta_{12} + 1}},
\\
&&{{\bf k}_3 \cdot {\bf k}_1 \over k_3 k_1}
=-{c \cos \theta_{12} +1\over \sqrt{c^2 + 2c \cos \theta_{12} + 1}}.
\end{eqnarray}
For convenience, we summarize the explicit expressions of 
$\alpha({\bf k}_i,{\bf k}_j)$ and $\gamma({\bf k}_i,{\bf k}_j)$. 
The above relations yield
\begin{eqnarray}
&&\alpha({\bf k}_1,{\bf k}_2) 
=1 + {(c^2 +1)\cos \theta_{12}\over 2c},
\\
&&\alpha({\bf k}_2,{\bf k}_3) 
=1 - {(2 c^2 + 2 c \cos \theta_{12} + 1)(c + \cos \theta_{12})\over 2c (c^2 + 2 c \cos \theta_{12} + 1)},
\\
&&\alpha({\bf k}_3,{\bf k}_1) 
=1- {(c^2 + 2 c \cos \theta_{12} + 2)(c \cos \theta_{12} + 1)\over 2 (c^2 + 2 c \cos \theta_{12} + 1)},
\\
&&\gamma({\bf k}_1,{\bf k}_2)=1 - \cos^2 \theta_{12},
\\
&&\gamma({\bf k}_2,{\bf k}_3)= {\sin^2 \theta_{12} \over c^2 + 2c \cos \theta_{12} + 1},
\\
&&\gamma({\bf k}_3,{\bf k}_1)
= {c^2 \sin^2 \theta_{12} \over c^2 + 2c \cos \theta_{12} + 1}.
\end{eqnarray}
Thus, $\alpha$ and $\gamma$ depend only on $c$ and $\theta_{12}$, which means
that $F_2(t,{\bf k}_i,{\bf k}_j)$ depends only on
$c$ and $\theta_{12}$, excepting $t$. It is trivial that 
$\alpha({\bf k}_1,{\bf k}_2)$ and $\gamma({\bf k}_1,{\bf k}_2)$
are invariant under the interchange between ${\bf k}_1$ and
${\bf k}_2$, or the replacement of $c$ with $1/c$. 
Note also that $\alpha({\bf k}_2,{\bf k}_3)$ and 
$\gamma({\bf k}_2,{\bf k}_3)$ are transformed into  
$\alpha({\bf k}_3,{\bf k}_1)$ and $\gamma({\bf k}_3,{\bf k}_1)$, 
respectively, by the replacement of $c$ with $1/c$.

\end{document}